\documentclass[preprint2]{aastex701}
\usepackage{multirow}

\begin{document}

\title{Emission-line Diagnostics at $z\gtrsim 2$: A Probe of the Ionizing Spectrum and $\alpha$ Enhancement Beyond Cosmic Noon}

\author[0000-0003-1249-6392]{Leonardo Clarke}
\affiliation{University of California, Los Angeles, 475 Portola Plaza, Los Angeles, CA, 90095, USA}
\email[show]{leoclarke@astro.ucla.edu}

\author[0000-0001-9489-3791]{Natalie Lam}
\altaffiliation{NSF Graduate Research Fellow}
\affiliation{University of California, Los Angeles, 475 Portola Plaza, Los Angeles, CA, 90095, USA}
\email{natalielamwy@astro.ucla.edu}

\author[0000-0003-3509-4855]{Alice E. Shapley}
\affiliation{University of California, Los Angeles, 475 Portola Plaza, Los Angeles, CA, 90095, USA}
\email{aes@astro.ucla.edu}

\author[0000-0001-8426-1141]{Michael W. Topping}
\affiliation{Steward Observatory, University of Arizona, 933 N Cherry Avenue, Tucson, AZ 85721, USA}
\email{michaeltopping@arizona.edu}

\author[0000-0003-2680-005X]{Gabriel B. Brammer}
\affiliation{Cosmic Dawn Center (DAWN), Denmark}
\affiliation{Niels Bohr Institute, University of Copenhagen, Jagtvej 128, DK-2200 Copenhagen N, Denmark}
\email{gabriel.brammer@nbi.ku.dk}

\author[0000-0003-4792-9119]{Ryan L. Sanders}
\affiliation{University of Kentucky, 506 Library Drive, Lexington, KY, 40506, USA}
\email{ryan.sanders@uky.edu}

\author[0000-0001-9687-4973]{Naveen A. Reddy}
\affiliation{Department of Physics \& Astronomy, University of California, Riverside, 900 University Avenue, Riverside, CA 92521, USA}
\email{naveenr@ucr.edu}

\author[0009-0002-6186-0293]{Shreya Karthikeyan}
\affiliation{University of California, Los Angeles, 475 Portola Plaza, Los Angeles, CA, 90095, USA}
\email{skarthik@g.ucla.edu}









\newcommand{\oiii}{[O\thinspace{\sc iii}]}
\newcommand{\neiii}{[Ne\thinspace{\sc iii}]}
\newcommand{\oii}{[O\thinspace{\sc ii}]}
\newcommand{\hii}{H\thinspace{\sc ii} }
\newcommand{\nii}{[N\thinspace{\sc ii}]}
\newcommand{\niii}{[N\thinspace{\sc iii}]}
\newcommand{\feii}{[Fe\thinspace{\sc ii}]}
\newcommand{\sii}{[S\thinspace{\sc ii}]}
\newcommand{\oi}{[O\thinspace{\sc i}]}

\begin{abstract}

We analyze several key rest-optical emission-line ratios in a sample of 828 galaxies as well as composite spectra from JADES DR3 in the range $1.4 < z < 7$. These emission-line ratios include: \oiii$\lambda5008$/H$\beta$, \nii$\lambda 6585$/H$\alpha$, \sii$\lambda\lambda 6718,6733$/H$\alpha$, \oi$\lambda 6302$/H$\alpha$, O32, R23, Ne3O2, and RO2Ne3. We find evidence for a harder ionizing spectrum at $z\sim 3.5$ compared to $z\sim 2$ at fixed gas-phase metallicity, resulting in a pronounced shift in the star-forming galaxy locus on the \nii/H$\alpha$ BPT diagram and the O32 vs. R23 diagram. At $z\gtrsim 3.5$, star-forming galaxies occupy a common locus, indicating that ISM ionizing conditions at fixed gas-phase metallicity do not evolve strongly at these early cosmic times. There is a connection between ISM ionizing conditions and the chemical abundance patterns (i.e., $\alpha$/Fe) in massive stars providing the ionizing radiation field. Therefore, the lack of evolution in ISM ionizing conditions at $z\gtrsim 3.5$, followed by evolution towards a softer ionizing spectrum at fixed nebular metallicity as cosmic time proceeds to $z\sim 2$ and lower redshift mirrors the chemical abundance patterns in Milky Way stars as a function of iron abundance.
Our results highlight the diagnostic power of emission-line diagrams in the era of JWST to further our understanding of the ISM conditions into the Epoch of Reionization.

\end{abstract}



\section{Introduction}

The near-infrared (NIR) sensitivity and continuous 1--5 $\rm \mu m$ coverage of JWST NIRCam \citep{2023PASP..135b8001R} and NIRSpec \citep{2022A&A...661A..80J} have revolutionized our ability to characterize the high-redshift galaxy population through measurements of rest-optical emission-line ratios \citep[e.g.,][]{2023A&A...677A.115C,2023ApJ...955...54S,2024MNRAS.529.3301T,2024ApJ...976..193R,2025ApJ...989L..55R,2025arXiv250708245T}. Prior to JWST, amassing large samples of star-forming galaxies with measurements of key emission line ratios such as \nii$\lambda 6585$/H$\alpha$ and \oiii$\lambda 5008$/H$\beta$ was only feasible out to $z\sim 2.6$ due to the wavelength-dependent transmission of the atmosphere in the NIR \citep[e.g.,][]{2014ApJ...795..165S,2015ApJ...801...88S,2017ApJ...836..164S,2019ApJ...881L..35S,2021MNRAS.502.2600R}. However, with large samples of JWST rest-optical galaxy spectra, it is now possible to push studies of the ISM out to higher redshifts and into the Epoch of Reionization.

A key goal in the era of JWST is to understand the conditions in \hii regions within high-redshift galaxies. It is now well-established that the typical conditions within star-forming (SF) \hii regions have evolved with cosmic time, reflected by a difference in the distribution of SF galaxies on the so-called ``BPT" diagrams \citep{1981PASP...93....5B, 1987ApJS...63..295V} at $z\sim 2$ compared to $z\sim 0$ galaxies and \hii regions. Notably, $z\sim 2$ galaxies exhibit increased \oiii$\lambda 5008$/H$\beta$ ratios at fixed \nii$\lambda 6585$/H$\alpha$ and \sii$\lambda\lambda 6718,6733$/H$\alpha$ relative to the locus of local \hii regions \citep[e.g.,][]{2014ApJ...795..165S,2015ApJ...801...88S,2017ApJ...836..164S,2017ApJ...835...88K,2019ApJS..241...10K,2019ApJ...881L..35S,2021MNRAS.502.2600R,2023ApJ...955...54S,2025ApJ...980..242S}. Much of this evolution can be attributed to a harder ionizing spectrum shape at fixed nebular metallicity, consistent with $\alpha$-enhanced abundance patterns \citep{2016ApJ...826..159S, 2020MNRAS.495.4430T, 2021MNRAS.502.2600R}. Other potential factors such as an evolving N/O abundance ratio, ionization parameter, and electron density may contribute to this evolution, but they are likely not the primary drivers of the observed evolution in these line ratios \citep{2016ApJ...816...23S,2020ApJ...888L..11S,2020MNRAS.495.4430T,2020MNRAS.499.1652T,2022MNRAS.512.2867H,2025ApJ...980..242S}. At $z\gtrsim 3$ however, the BPT line ratios have only been measured in relatively small (i.e., $\lesssim$100) samples of SF galaxies \citep[e.g.,][]{2023ApJ...955...54S,2023A&A...677A.115C,2025ApJ...980..242S}.

Another common diagram is the O32 (\oiii/\oii) vs. R23 ((\oiii+\oii)/H$\beta$) diagram. R23 is a common nebular metallicity indicator, while O32 is correlated with the ionization parameter, making this diagram a useful tracer of metallicity and ionization conditions \citep[e.g.,][]{2001ApJ...556..121K, 2020MNRAS.491..944C, 2025arXiv250810099S}. For galaxies at $z> 9.6$, where \oiii$\lambda 5008$ redshifts beyond the wavelength coverage of NIRSpec, \citet{2023A&A...677A..88B} proposed the use of a neon-based diagram, replacing \oiii$\lambda 5008$ with \neiii$\lambda 3870$ and H$\beta$ with H$\delta$. Because the abundances of neon and oxygen trace each other closely \citep[e.g.,][]{2014ApJ...780..100L}, and Ne$^{2+}$ and O$^{2+}$ follow similar spatial distributions in \hii regions due to their similar ionization energies, \neiii\ emission can be used as a proxy for \oiii, and thus serves a useful role in metallicity and ionization parameter diagnostics \citep[e.g.,][]{2007MNRAS.381..125P}. Similar to the BPT diagrams, studies of galaxies using these diagrams are limited in sample size (i.e., $\lesssim$120) at $z\gtrsim 3$ \citep[e.g.,][]{2023A&A...677A.115C,2023ApJ...955...54S,2025ApJ...980..242S}.

In this study, we present a sample of 828 galaxies from DR3 of the JADES survey in the range $1.4<z<7$, largely representative of the SF galaxy population above $10^{8.5}\rm\ M_\odot$ as discussed in \citet{2025arXiv251006681C}. We analyze the distributions of these galaxies on the BPT diagrams as well as the aforementioned O32 vs. R23 and neon-based diagrams. Using these diagrams, we illustrate the redshift evolution of the ISM conditions in star-forming galaxies.

We define the names of the following emission-line ratios as follows:

\begin{enumerate}
    \item O32$^*$: $\rm \log_{10}\left (\frac{[O\thinspace III] \lambda5008}{[O\thinspace II]\lambda\lambda 3727, 3730} \right)$
    \item R23$^*$: $\rm \log_{10}\left (\frac{[O\thinspace III]\lambda\lambda 4960,5008 + [O\thinspace II]\lambda \lambda 3727,3730}{H\beta}\right)$
    \item Ne3O2: $\rm \log_{10}\left (\frac{[Ne\thinspace III]\lambda 3870}{[O\thinspace II]\lambda\lambda3727,3730}\right)$  
    \item RO2Ne3: $\rm \log_{10}\left (\frac{[Ne\thinspace III]\lambda3870+[O\thinspace II]\lambda\lambda3727,3730}{H\delta}\right)$
\end{enumerate}

where line centroids are quoted in vacuum wavelengths. Line ratios marked with an asterisk (*) are corrected for dust attenuation due to the significant wavelength separation of the component lines. In Section \ref{sec:observations}, we describe the observations, emission-line measurements, and derivation of physical parameters. In Section \ref{sec:results}, we present measurements of several key emission-line diagrams. In Section \ref{sec:discussion}, we discuss and summarize our results. Throughout this Letter, we assume $H_0 = 70\ \rm km\ s^{-1}\  Mpc^{-1}$, $\Omega_m = 0.3$, $\Omega_\Lambda = 0.7$, and a \citet{2003PASP..115..763C} initial-mass function.


\section{Observations and Measurements} \label{sec:observations}

The observations and measurements on which this work is based are described in \citet{2025arXiv251006681C}. We provide a brief overview in the following sub-sections.

\subsection{NIRCam and NIRSpec Observations}

The catalog of JWST/NIRCam and HST photometric measurements used to determine stellar population properties can be found on the JADES MAST page\footnote{https://archive.stsci.edu/hlsp/jades}. Specifically, we used the \texttt{KRON\_CONV} table extension, which consists of photometry measured in Kron apertures, where the point-spread function of all images was homogenized to match that of the F444W filter. This catalog compiles the broad- and medium-band observations from the JADES NIRCam programs in GOODS-N and GOODS-S \citep{2023ApJS..269...16R,2023arXiv231012340E}, F444W photometry from FRESCO \citep{2023MNRAS.525.2864O}, and medium-band photometry from JEMS \citep{2023ApJS..268...64W}. 

The JWST/NIRSpec observations analyzed in this study come from data release 3 (DR3) of the JADES program \citep{2023arXiv230602465E,2024A&A...690A.288B,2025ApJS..277....4D} and were downloaded from the Dawn JWST Archive (DJA)\footnote{https://dawn-cph.github.io/dja/}. Targets were observed using both the $R\sim 100$ prism and $R\sim 1000$ grating (G140M/F100LP, G235M/F170LP, and G395M/F290LP) dispersers. We describe the flux calibration of these spectra in section \ref{sec:fluxcal_emline}.

\subsection{Stellar population fitting}

The stellar population parameters (e.g., stellar mass, star-formation history, stellar dust attenuation) were obtained through SED fitting with the {\sc prospector} code \citep{2021ApJS..254...22J}. The details of the SED fitting are described in \citet{2025arXiv251006681C}. In brief, we simultaneously fit SED models to the NIRCam+HST photometry and the flux-calibratied prism spectra. Additionally, we adopted a non-parametric star-formation history (SFH) with a Student-t distribution as a continuity prior, dividing the SFH into eight time bins, with the two most recent bins spanning 3 Myr and 10 Myr. The remaining six bins were equally logarithmically spaced, with the earliest bin starting at $z=20$. The best-fit stellar and nebular continuum SED models served as the basis for modeling the continuum emission during emission-line fitting.

\subsection{Flux calibration and line-flux measurement}\label{sec:fluxcal_emline}

As described in \citet{2025arXiv251006681C}, we scaled the prism spectra with a wavelength-dependent polynomial to match the flux calibration to the avaliable NIRCam and HST photometry. Because the traces of many of the grating spectra overlap each other on the NIRSpec detector \citep{2024A&A...690A.288B,2025ApJS..277....4D}, we did not scale the $R\sim 1000$ grating spectra to match the photometry, and we instead used a custom approach to ensure the accuracy of the flux ratios of emission lines that lie in different gratings. For line flux ratios in individual targets, the fitting procedure is described in \citet{2025arXiv251006681C}, whereby lines were fit in the prism and grating spectra simultaneously, leveraging both the accurate flux calibration of the prism and the resolution of the gratings. Emission-line fluxes  were corrected for dust attenuation using the methodology of \citet{2025arXiv251006681C}. 

\begin{figure*}
    \centering
    \includegraphics[width=\linewidth]{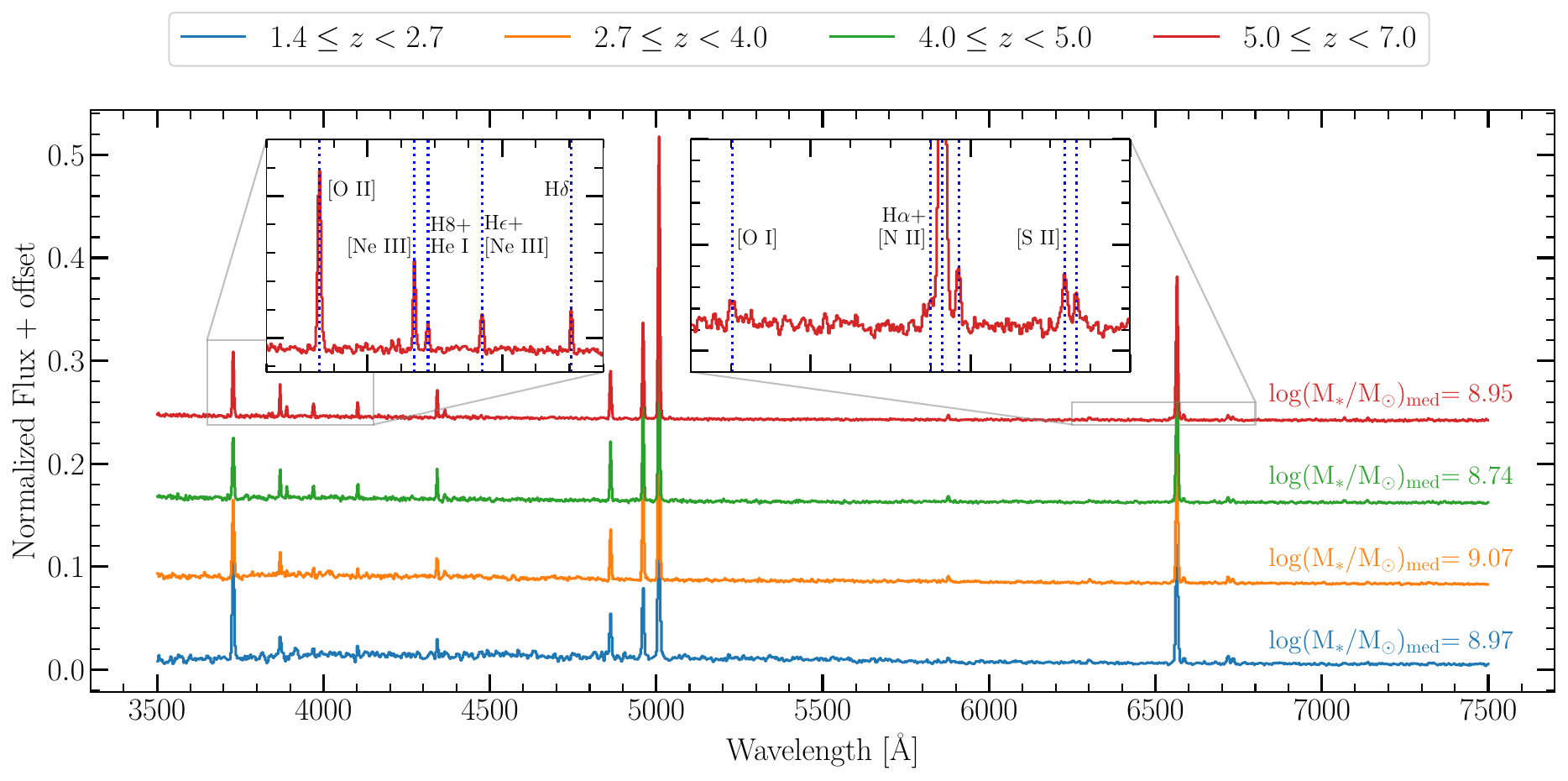}
    \caption{Stacked composite spectra in each of the four redshift bins examined in this study. As labeled, each stack is composed of galaxies with median stellar masses of $\sim$$10^9\ \rm M_\odot$. The insets showcase several faint emission lines analyzed in this study.}
    \label{fig:stacked_specs}
\end{figure*}

\subsection{Composite spectra}\label{sec:composite}

In addition to single-galaxy line flux ratio measurements, we also present measurements from stacked composite spectra, generated from the grating spectra. As described in detail in \citet{2026arXiv260311338K}, we implemented a robust cross-grating flux calibration by comparing the fluxes of strong emission lines in the gratings to the same lines measured in the prism, and scaling the grating spectra accordingly to match the flux-calibrated prism-based measurements. For those objects that did not have $>$3$\sigma$ detections of emission lines in both the grating and the prism spectra, we did not apply any multiplicative scaling factor. In the following subsection, we describe the selection criteria applied to the sample that are included in the stacked spectra.

\subsubsection{Sample Selection}\label{sec:sample_selection}

From the parent sample of 4,086 galaxies in DR3 of JADES, we selected galaxies in the range $1.4 < z < 7$ for which the NIRSpec observations did not suffer from microshutter failures \citep[see][]{2025ApJS..277....4D}, for which there were $>$5$\sigma$ JWST/NIRCam and/or HST imaging detections, and that could be robustly flux calibrated (see Appendix B of \citealt{2025arXiv251006681C}). These criteria resulted in a sample of 1,245 galaxies. From this sample, we further removed galaxies with specific SFR $<10^{-11}\ \rm yr^{-1}$, and required a $\geq$3$\sigma$ detection of H$\alpha$ in both the grating and prism observations of a given object, removing another 210 objects. We additionally imposed criteria to flag AGN-dominated objects, removing 32 targets for which \nii/H$\alpha >= 0.5$ and 9 objects with significant broad Balmer line profiles. Finally, to ensure a robust flux calibration and measurement of \oii$ \lambda\lambda 3727,3730$, we removed 10 objects in which the \oii\ lines were significantly blended with the Balmer break in the prism, a known issue that inflates the uncertainty of prism-based \oii\ flux estimates \citep[see][]{2025ApJS..277....4D}. Collectively, these criteria resulted in a sample of 984 galaxies to which we apply our binning scheme.

We binned the galaxies by redshift and stellar mass in four different sets, where all spectra within a given set cover the following lines:

\begin{enumerate}
    \item BPT Set: H$\beta$, \oiii, H$\alpha$, \nii, and \sii
    \item \oi\ Set: H$\beta$, \oiii, \oi, and H$\alpha$
    \item O32R23 Set: \oii, H$\beta$, \oiii, and H$\alpha$
    \item \neiii\  Set: \oii, \neiii, H$\delta$, and H$\alpha$
\end{enumerate}

The BPT and \oi\ Sets include coverage of the BPT diagram lines, while the O32R23 and \neiii\ Sets cover the lines used in the diagrams discussed in section \ref{sec:o32_r23}. Imposing the additional criteria of requiring line coverage in each of the stack sets results in 693, 693, 608, and 684 objects in the BPT, \oi, O32R23, and RO2Ne3 stack sets, respectively. Taken together, across all four stack sets, there are 828 unique objects, comprising the total number of objects included in this analysis. The objects in these stacks are consistent with the star-forming main sequence (SFMS) determined by \citet{2025MNRAS.544.4551S} at $z>3$ and by \citet{2025arXiv251006681C} at $1.4<z<2.7$, with the median of the SFRs of the sample above $10^{8.5}\ \rm M_\odot$ agreeing to within 0.08 dex of the SFMS across redshift bins. In all sets, the lowest-mass bin contains galaxies with $\log_{10}\left(\mathrm{M}_{\ast}/\mathrm{M}_{\odot}\right) < 8.5$, the mass below which the sample is representative \citep{2025arXiv251006681C}. Galaxies with $\log_{10}\left(\mathrm{M}_{\ast}/\mathrm{M}_{\odot}\right)\geq 8.5$ are binned such that each stack contains a similar number of objects. Because objects move between mass bins during the Monte Carlo simulations, the quoted numbers of objects per bin throughout this Letter represent the initial number of objects assigned to a bin prior to perturbing the stellar masses.

\subsubsection{Stacking Procedure}

To produce the stacked spectrum for each set of galaxies, we first de-redshifted and normalized each spectrum by the H$\alpha$ luminosity and resampled the spectra onto a common wavelength sampling grid using the {\sc SpectRes} package \citep{2017arXiv170505165C}. The wavelength sampling of a given stacked spectrum was set to the median value of the wavelength samplings of all spectra that comprised that stack. After resampling, we calculated the median flux density $f_\lambda$ at each wavelength to obtain a median stacked spectrum. We performed Monte Carlo simulations, repeating this stacking process 100 times, perturbing the stellar masses by their respective uncertainties to allow objects to shift between stellar mass bins, bootstrap resampling within each bin to account for sample variance, and perturbing each spectrum by its associated uncertainty at each wavelength. Following these simulations, we calculated the median and standard deviation of $f_\lambda$ values at each wavelength, and we adopted these as the final composite flux and error spectra, respectively. Line ratios in the O32R23 composite Set are corrected for dust using the stacked H$\alpha$/H$\beta$ Balmer decrement, adopting an intrinsic ratio of 2.79, consistent with an electron temperature of 15,000 K and electron density of 100 $\rm cm^{-3}$, as calculated using {\sc PyNeb} \citep{2015A&A...573A..42L}. Additionally, we assume a \citet{1989ApJ...345..245C} dust law, as in \citet{2025arXiv251006681C} and \citet{2026arXiv260311338K}. We showcase four stacked spectra from the BPT Set in Figure \ref{fig:stacked_specs} and all stacked line ratio measurements in Tables \ref{tab:line_ratios2} and \ref{tab:line_ratios}.

\section{Results} \label{sec:results}

\subsection{BPT Line Ratios}

\begin{figure*}
    \centering
    \includegraphics[width=\textwidth]{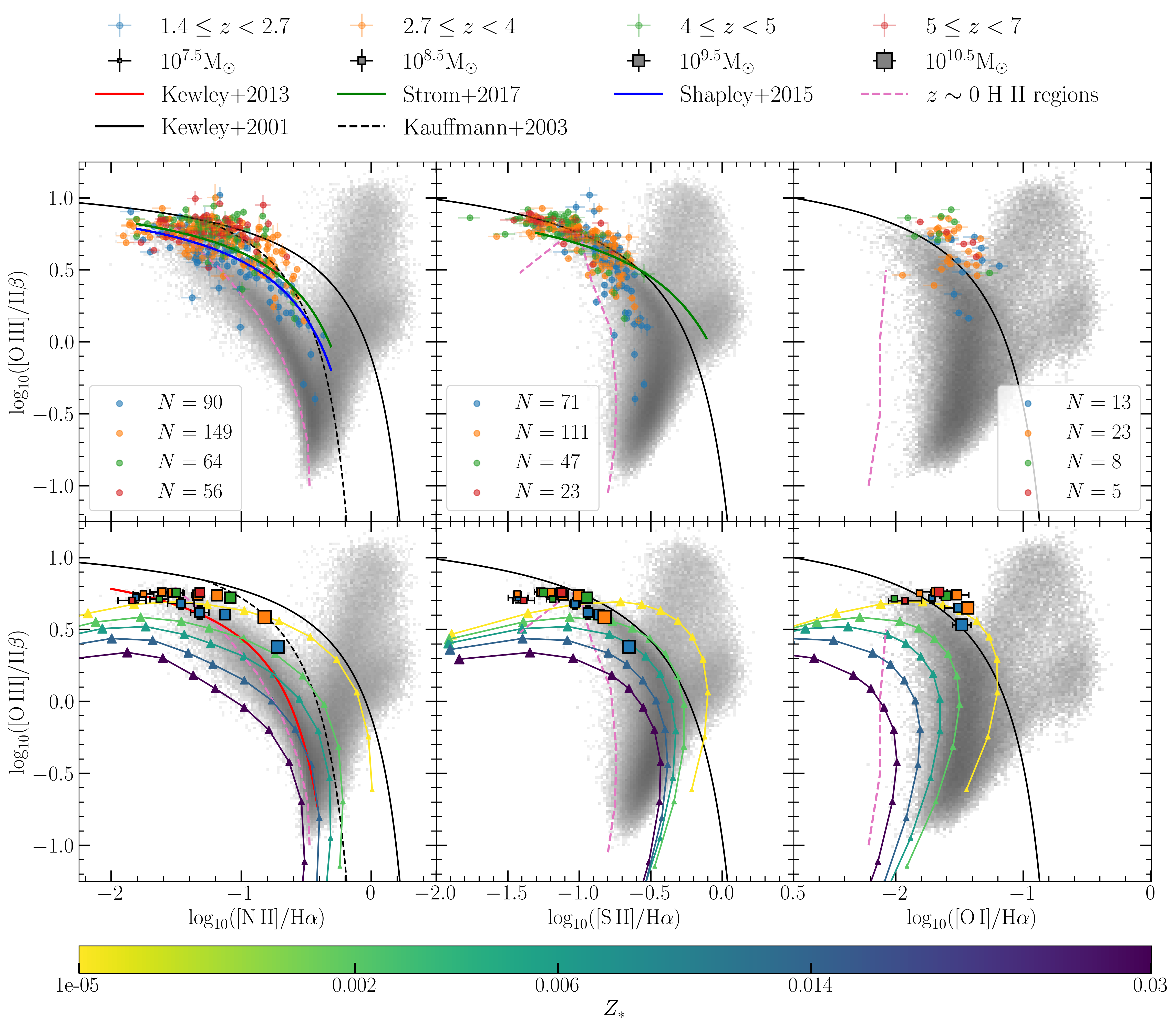}
    \caption{BPT diagrams for the JADES DR3 sample. The boundaries separating SF galaxies from AGN from \citet{2001ApJ...556..121K} and \citet{2003MNRAS.346.1055K} are shown as solid and dashed black lines, respectively. The gray 2D histogram shows the distribution of $z \sim 0$ galaxies from SDSS \citep[DR7;][]{2009ApJS..182..543A}. The pink dashed curve shows the binned median sequence of local \hii regions from the CHAOS survey \citep[e.g.,][]{2015ApJ...806...16B} and a compilation by \citet{2016MNRAS.457.3678P}. The \nii\ and \oi\ diagrams bin by \oiii/H$\beta$, while the \sii\ diagram bins by electron temperature. Cloudy curves from \citet{2020ApJ...902L..16J} are overlaid and color-coded by stellar metallicity. The sizes of the triangles on the model curves correspond to the ionization parameter, where larger symbols represent a larger ionization parameter}. Emission-line sequences from \citet{2013ApJ...774..100K} ($z\sim 0$; lower \nii\ panel), \citet{2015ApJ...801...88S} ($z\sim 2$; upper \nii\ panel), and \citet{2017ApJ...836..164S} ($z\sim 2$; upper \nii\ and \sii\ panels) are shown for reference.
    \label{fig:bpt}
\end{figure*}

\begin{deluxetable}{c|ccccc|cccc}
\label{tab:line_ratios2}
\caption{Emission-line ratios for stacked measurements shown in Figure \ref{fig:bpt}}
\tablehead{\multirow{2}{*}{$z_{\rm range}$} & \multicolumn{5}{c|}{BPT Set} & \multicolumn{4}{c}{\oi\ Set} \\ 
 & \colhead{N} & \colhead{$\rm \log(\frac{M_*}{M_\odot})$} & \colhead{[O\thinspace{\sc iii}]/H$\beta$} & \colhead{[N\thinspace{\sc ii}]/H$\alpha$} & \colhead{[S\thinspace{\sc ii}]/H$\alpha$} \vline & \colhead{N} & \colhead{$\rm \log(\frac{M_*}{M_\odot})$} & \colhead{[O\thinspace{\sc iii}]/H$\beta$} & \colhead{[O\thinspace{\sc i}]/H$\alpha$}}
\startdata
\multirow{5}{*}{$1.4-2.7$} & 50 & $ 8.18^{+0.03}_{-0.03}$ & $ 0.72^{+0.04}_{-0.03}$ & $ -1.81^{+0.01}_{-0.01}$ & $ -1.43^{+0.07}_{-0.07}$ &  50 & $ 8.18^{+0.04}_{-0.03}$ & $ 0.72^{+0.04}_{-0.04}$ & $ -1.72^{+0.01}_{-0.01}$ \\
 &  39 & $ 8.67^{+0.03}_{-0.02}$ & $ 0.68^{+0.04}_{-0.04}$ & $ -1.46^{+0.10}_{-0.11}$ & $ -1.03^{+0.04}_{-0.04}$ &  77 & $ 8.82^{+0.03}_{-0.02}$ & $ 0.65^{+0.03}_{-0.03}$ & $ -1.51^{+0.09}_{-0.13}$ \\
 &  38 & $ 8.97^{+0.03}_{-0.02}$ & $ 0.62^{+0.05}_{-0.04}$ & $ -1.32^{+0.07}_{-0.08}$ & $ -0.94^{+0.03}_{-0.03}$ &  76 & $ 9.46^{+0.03}_{-0.02}$ & $ 0.53^{+0.03}_{-0.03}$ & $ -1.48^{+0.06}_{-0.08}$ \\
 &  38 & $ 9.25^{+0.02}_{-0.02}$ & $ 0.60^{+0.03}_{-0.04}$ & $ -1.13^{+0.03}_{-0.04}$ & $ -0.86^{+0.03}_{-0.04}$ & --  & --  & --  & -- \\
 &  38 & $ 9.78^{+0.03}_{-0.02}$ & $ 0.38^{+0.03}_{-0.03}$ & $ -0.72^{+0.02}_{-0.02}$ & $ -0.65^{+0.02}_{-0.02}$ & --  & --  & --  & -- \\
\hline
\multirow{6}{*}{$2.7-4.0$} & 82 & $ 8.16^{+0.04}_{-0.04}$ & $ 0.75^{+0.02}_{-0.02}$ & $ -1.75^{+0.07}_{-0.08}$ & $ -1.43^{+0.03}_{-0.03}$ &  82 & $ 8.15^{+0.04}_{-0.04}$ & $ 0.75^{+0.02}_{-0.02}$ & $ -1.81^{+0.01}_{-0.01}$ \\
 &  38 & $ 8.63^{+0.03}_{-0.02}$ & $ 0.76^{+0.02}_{-0.02}$ & $ -1.61^{+0.08}_{-0.11}$ & $ -1.27^{+0.04}_{-0.05}$ &  63 & $ 8.72^{+0.02}_{-0.03}$ & $ 0.76^{+0.02}_{-0.02}$ & $ -1.69^{+0.11}_{-0.17}$ \\
 &  38 & $ 8.88^{+0.02}_{-0.03}$ & $ 0.76^{+0.02}_{-0.02}$ & $ -1.53^{+0.08}_{-0.10}$ & $ -1.20^{+0.05}_{-0.05}$ &  63 & $ 9.07^{+0.02}_{-0.02}$ & $ 0.74^{+0.02}_{-0.01}$ & $ -1.52^{+0.08}_{-0.11}$ \\
 &  38 & $ 9.07^{+0.02}_{-0.02}$ & $ 0.74^{+0.02}_{-0.02}$ & $ -1.33^{+0.04}_{-0.05}$ & $ -1.11^{+0.02}_{-0.03}$ &  62 & $ 9.67^{+0.02}_{-0.03}$ & $ 0.65^{+0.01}_{-0.01}$ & $ -1.44^{+0.03}_{-0.04}$ \\
 &  37 & $ 9.35^{+0.02}_{-0.02}$ & $ 0.74^{+0.02}_{-0.01}$ & $ -1.19^{+0.02}_{-0.02}$ & $ -1.00^{+0.02}_{-0.03}$ & --  & --  & --  & -- \\
 &  37 & $ 9.92^{+0.02}_{-0.03}$ & $ 0.59^{+0.01}_{-0.01}$ & $ -0.82^{+0.01}_{-0.01}$ & $ -0.82^{+0.02}_{-0.02}$ & --  & --  & --  & -- \\
\hline
\multirow{3}{*}{$4.0-5.0$} & 34 & $ 8.16^{+0.06}_{-0.06}$ & $ 0.71^{+0.02}_{-0.02}$ & $ -1.63^{+0.07}_{-0.07}$ & $ -1.19^{+0.04}_{-0.05}$ &  34 & $ 8.16^{+0.04}_{-0.06}$ & $ 0.71^{+0.02}_{-0.02}$ & $ -2.01^{+0.01}_{-0.01}$ \\
 &  37 & $ 8.74^{+0.04}_{-0.03}$ & $ 0.76^{+0.02}_{-0.02}$ & $ -1.50^{+0.05}_{-0.05}$ & $ -1.25^{+0.02}_{-0.02}$ &  73 & $ 8.96^{+0.05}_{-0.04}$ & $ 0.73^{+0.01}_{-0.01}$ & $ -1.61^{+0.04}_{-0.05}$ \\
 &  36 & $ 9.35^{+0.05}_{-0.03}$ & $ 0.72^{+0.01}_{-0.01}$ & $ -1.08^{+0.02}_{-0.02}$ & $ -0.95^{+0.02}_{-0.02}$ & --  & --  & --  & -- \\
\hline
\multirow{2}{*}{$5.0-7.0$} & 62 & $ 8.16^{+0.04}_{-0.04}$ & $ 0.70^{+0.01}_{-0.01}$ & $ -1.84^{+0.10}_{-0.12}$ & $ -1.39^{+0.07}_{-0.09}$ &  62 & $ 8.15^{+0.04}_{-0.04}$ & $ 0.70^{+0.01}_{-0.01}$ & $ -1.93^{+0.10}_{-0.16}$ \\
 &  51 & $ 8.95^{+0.05}_{-0.05}$ & $ 0.76^{+0.01}_{-0.01}$ & $ -1.32^{+0.03}_{-0.03}$ & $ -1.12^{+0.02}_{-0.02}$ &  51 & $ 8.95^{+0.04}_{-0.06}$ & $ 0.76^{+0.01}_{-0.01}$ & $ -1.66^{+0.06}_{-0.07}$ \\
\hline
\enddata
\end{deluxetable}

In Figure \ref{fig:bpt}, we show our sample on the BPT diagrams. The gray 2D histograms show the distribution of galaxies at $z\sim 0$ from the Sloan Digital Sky Survey \citep[SDSS][]{2009ApJS..182..543A}, and the SF galaxy boundaries from \citet{2001ApJ...556..121K} and \citet{2003MNRAS.346.1055K} are shown as solid and dashed black curves, respectively. The colored points show emission-line ratios in individual galaxies, color-coded by the redshift range. From the initial spectroscopic sample of 1,245 objects, the galaxies that are plotted have 5$\sigma$ detections each of the component lines in each diagram. After removing AGN (based on the same criteria applied to the stacked spectra), this selection results in a sample of 359, 252, and 49 individual galaxies plotted on the \nii, \sii, and \oi\ BPT diagrams, respectively. These individual objects are also roughly consistent with the \citet{2025MNRAS.544.4551S} and \citet{2025arXiv251006681C} SFMS determinations, with the median offset agreeing within 0.05 dex across redshift. As in \citet{2025arXiv251006681C}, we divide the sample into four redshift ranges: $1.4\leq z < 2.7$, $2.7 \leq z < 4$, $4\leq z < 5$, and $5\leq z < 7$. As a short-hand, we refer to these ranges as $z\sim 2$, $z\sim 3.5$, $z\sim 4.5$, and $z\sim 6$, respectively. The pink dashed curves show the binned median line ratios on the BPT diagrams from local \hii regions from the CHAOS survey \citep{2015ApJ...806...16B} and a compilation by \citet{2016MNRAS.457.3678P}. We also show photoionization model curves from Cloudy \citep{2017RMxAA..53..385F} for various stellar metallicities, described in \citet{2020ApJ...902L..16J}. Though the normalizations of these curves with respect to metallicity may not overlap precisely with the measured emission-line ratios, our primary focus is the differential between each curve, informing us as to how differences in stellar metallicity translate to differences in line ratios. For all emission-line ratios shown, we require a $>$5$\sigma$ detection of each component line. We additionally remove any AGN-dominated sources by requiring that \nii/H$\alpha < 0.5$, and that there are no prominent broad Balmer components.

In the \nii/H$\alpha$ BPT diagram, the SF locus at $z\sim 2$ is offset relative to the $z\sim 0$ sequence defined by \citet{2013ApJ...774..100K} by a value of $ 0.10\pm 0.03$ dex toward higher \oiii/H$\beta$ at a fixed value of log(\nii/H$\alpha)=-1$. At $z\sim 3.5$, this offset is larger, with a value of $ 0.22\pm 0.01$ dex. Because the $z  \sim 6$ stacks do not extend higher than $ -1.2$ in log(\nii/H$\alpha$), it is more useful to calculate an offset at fixed log(\nii/H$\alpha)=-1.2$ for the higher-$z$ sequences. In this case, the offsets of the $z\sim  3.5$, $z\sim  4.5$, and $z\sim  6$ sequences are $ 0.18\pm0.01$, $ 0.18\pm0.01$, and $ 0.20\pm0.01$ dex relative to the \citet{2013ApJ...774..100K} sequence, showing no strong time evolution at higher redshifts. In the \sii/H$\alpha$ diagram, at a fixed value of log(\sii/H$\alpha)=-1$, the stacked line ratios at $z\sim 3.5-6$ agree well with the local median SDSS sequence, evolving toward lower \oiii/H$\beta$ at fixed \sii/H$\alpha$ at $z\sim 2$, reaching an offset of $ -0.06\pm0.04$ dex relative to the median $z\sim 0$ locus by $z\sim 2$. Relative to the local \hii regions, the SF sequences at all redshifts exhibit elevated \oiii/H$\beta$ at both fixed \nii/H$\alpha$ and fixed \sii/H$\alpha$.

We also show our individual galaxy and stacked line measurements on the \oi/H$\alpha$ BPT diagram. Due to the intrinsic faintness of the \oi $\lambda6302$ line, the distribution of galaxies at $z>1$ on this diagram has not been widely explored \citep[e.g.,][]{2023ApJ...957...81C,2023A&A...677A.115C,2023ApJ...955...54S,2025ApJ...980..242S,2025ApJ...989L..55R}. As in the \sii/H$\alpha$ diagram, SF galaxies occupy an overlapping sequence on the \oi/H$\alpha$ diagram at $z\gtrsim 3.5$, with median evolution of $ 0.09\pm0.08$ dex toward higher \oiii/H$\beta$ at fixed \oi/H$\alpha$ from $z\sim 2$ to $z\sim 3.5$.

\begin{figure}[ht!]
    \centering
    \includegraphics[width=8.5cm]{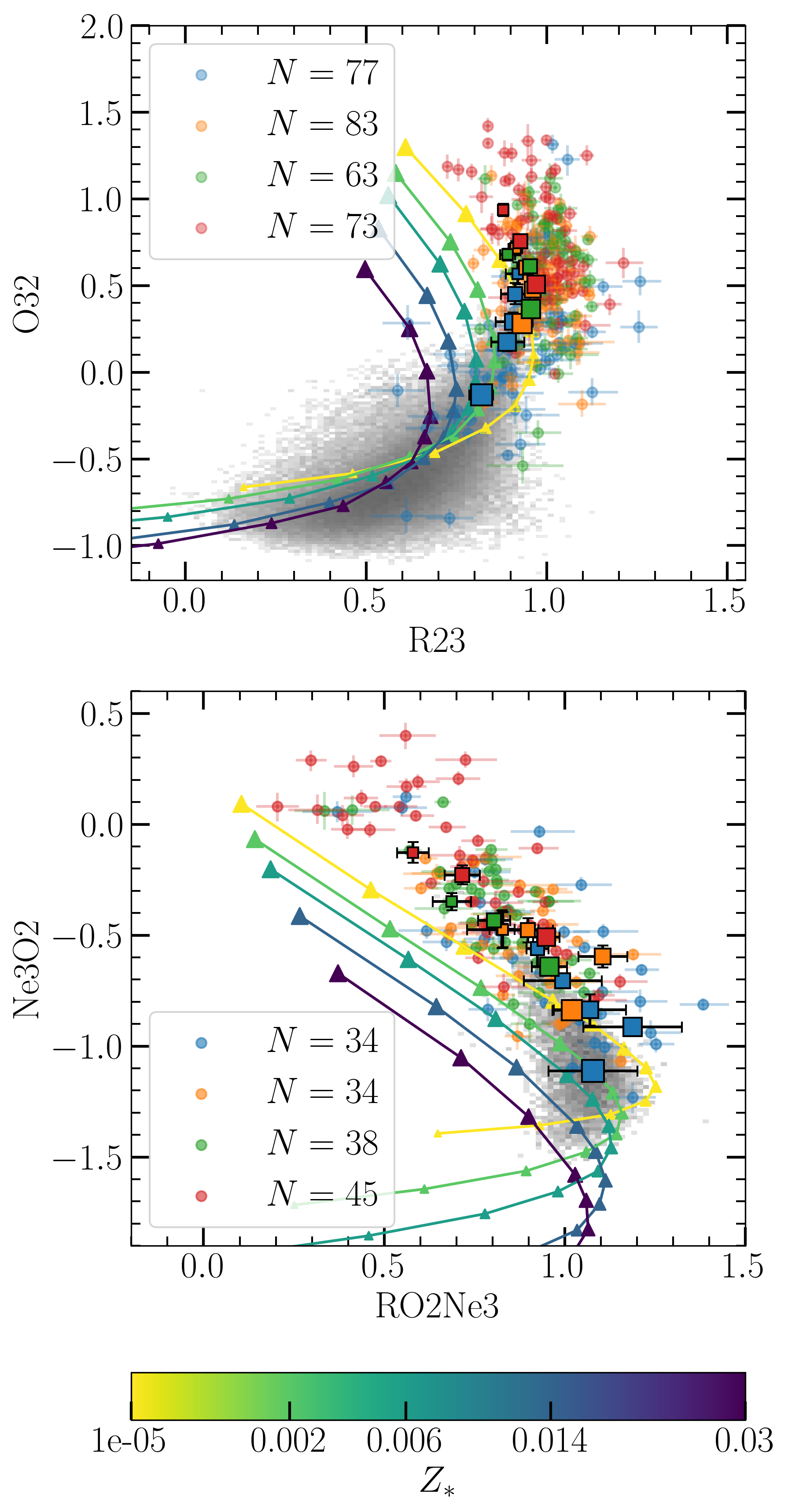}
    \caption{{\it Top:} O32 vs. R23 diagram. {\it Bottom:} Ne3O2 vs. RO2Ne3 diagram. The gray 2D histogram shows $z\sim 0$ galaxies, where AGN are removed based on the \citet{2003MNRAS.346.1055K} criterion. The color coding follows the legend in Figure \ref{fig:bpt}.}
    \label{fig:o32_r23}
\end{figure}

\subsection{Excitation and metallicity diagrams}\label{sec:o32_r23}

In addition to the BPT diagrams, we show the distribution of our sample on the O32 vs. R23 and Ne3O2 vs. RO2Ne3 excitation/metallicity diagrams. From the original 1,245 objects at $ 1.4 < z < 2.7$, we have plotted a subset of individual galaxy measurements on these diagrams, requiring 5$\sigma$ detections of all component lines. We also removed AGN-dominated objects with broad Balmer lines or \nii/H$\alpha > 0.5$ and removed unreliable \oii\ measurements, as described in Section \ref{sec:sample_selection}. These criteria result in 296 and 151 individual galaxies plotted in the top and bottom panels of Figure \ref{fig:o32_r23}, respectively. The individual galaxies shown in these diagrams agree with the \citet{2025MNRAS.544.4551S} and \citet{2025arXiv251006681C} SFMS within 0.14 dex. As in previous work \citep[e.g.,][]{2016ApJ...816...23S,2023ApJ...950L...1S}, we observe a shift toward increasing O32 with increasing redshift at fixed stellar mass (top panel of Figure \ref{fig:o32_r23}). We also observe a slight evolution toward higher R23 at fixed O32 from $z\sim 2$ to $z \sim 3.5$, while the SF locus at $z\gtrsim 3.5$ exhibits little time evolution on the O32 vs. R23 diagram. 

\begin{deluxetable}{c|cccc|cccc}
\label{tab:line_ratios}
\caption{Emission-line ratios for stacked measurements shown in Figure \ref{fig:o32_r23}}
\tablehead{\multirow{2}{*}{$z_{\rm range}$} & \multicolumn{4}{c|}{O32R23 Set} & \multicolumn{4}{c}{\neiii\ Set} \\ 
 & \colhead{N} & \colhead{$\rm \log(\frac{M_*}{M_\odot})$} & \colhead{O32} & \colhead{R23} \vline & \colhead{N} & \colhead{$\rm \log(\frac{M_*}{M_\odot})$} & \colhead{Ne3O2} & \colhead{RO2Ne3}}
\startdata
\multirow{6}{*}{$1.4-2.7$} & 49 & $ 8.17^{+0.04}_{-0.03}$ & $ 0.57^{+0.06}_{-0.06}$ & $ 0.92^{+0.04}_{-0.03}$ &  67 & $ 8.20^{+0.04}_{-0.02}$ & $ -0.47^{+0.08}_{-0.09}$ & $ 0.83^{+0.04}_{-0.03}$\\
 &  39 & $ 8.68^{+0.02}_{-0.02}$ & $ 0.45^{+0.06}_{-0.06}$ & $ 0.92^{+0.04}_{-0.04}$ &  43 & $ 8.66^{+0.03}_{-0.03}$ & $ -0.56^{+0.07}_{-0.09}$ & $ 0.92^{+0.03}_{-0.03}$\\
 &  38 & $ 8.98^{+0.03}_{-0.03}$ & $ 0.28^{+0.03}_{-0.05}$ & $ 0.90^{+0.05}_{-0.05}$ &  43 & $ 8.90^{+0.02}_{-0.02}$ & $ -0.71^{+0.03}_{-0.03}$ & $ 0.99^{+0.12}_{-0.10}$\\
 &  38 & $ 9.26^{+0.02}_{-0.02}$ & $ 0.17^{+0.05}_{-0.05}$ & $ 0.89^{+0.05}_{-0.04}$ &  43 & $ 9.15^{+0.02}_{-0.02}$ & $ -0.84^{+0.06}_{-0.08}$ & $ 1.07^{+0.11}_{-0.09}$\\
 &  38 & $ 9.79^{+0.03}_{-0.02}$ & $ -0.13^{+0.03}_{-0.04}$ & $ 0.82^{+0.03}_{-0.03}$ &  43 & $ 9.43^{+0.02}_{-0.03}$ & $ -0.91^{+0.02}_{-0.02}$ & $ 1.19^{+0.16}_{-0.12}$\\
 &  --  & --  & --  & --  &  43 & $ 9.95^{+0.02}_{-0.03}$ & $ -1.11^{+0.02}_{-0.02}$ & $ 1.08^{+0.14}_{-0.11}$\\
\hline
\multirow{4}{*}{$2.7-4.0$} & 51 & $ 8.17^{+0.06}_{-0.04}$ & $ 0.71^{+0.04}_{-0.04}$ & $ 0.91^{+0.02}_{-0.02}$ &  51 & $ 8.17^{+0.04}_{-0.04}$ & $ -0.48^{+0.07}_{-0.08}$ & $ 0.83^{+0.11}_{-0.09}$\\
 &  44 & $ 8.72^{+0.03}_{-0.04}$ & $ 0.60^{+0.03}_{-0.03}$ & $ 0.94^{+0.02}_{-0.02}$ &  42 & $ 8.71^{+0.03}_{-0.03}$ & $ -0.48^{+0.06}_{-0.05}$ & $ 0.90^{+0.07}_{-0.07}$\\
 &  44 & $ 9.04^{+0.02}_{-0.02}$ & $ 0.48^{+0.03}_{-0.03}$ & $ 0.96^{+0.02}_{-0.02}$ &  42 & $ 9.05^{+0.02}_{-0.02}$ & $ -0.60^{+0.05}_{-0.05}$ & $ 1.11^{+0.07}_{-0.06}$\\
 &  43 & $ 9.57^{+0.04}_{-0.03}$ & $ 0.28^{+0.02}_{-0.02}$ & $ 0.93^{+0.01}_{-0.02}$ &  42 & $ 9.61^{+0.05}_{-0.04}$ & $ -0.84^{+0.04}_{-0.05}$ & $ 1.02^{+0.06}_{-0.05}$\\
\hline
\multirow{3}{*}{$4.0-5.0$} & 33 & $ 8.16^{+0.05}_{-0.05}$ & $ 0.68^{+0.03}_{-0.04}$ & $ 0.89^{+0.02}_{-0.02}$ &  34 & $ 8.15^{+0.05}_{-0.06}$ & $ -0.35^{+0.04}_{-0.04}$ & $ 0.69^{+0.06}_{-0.05}$\\
 &  37 & $ 8.74^{+0.03}_{-0.04}$ & $ 0.61^{+0.03}_{-0.03}$ & $ 0.95^{+0.02}_{-0.02}$ &  40 & $ 8.74^{+0.03}_{-0.03}$ & $ -0.43^{+0.03}_{-0.03}$ & $ 0.80^{+0.04}_{-0.04}$\\
 &  36 & $ 9.35^{+0.04}_{-0.03}$ & $ 0.37^{+0.02}_{-0.02}$ & $ 0.96^{+0.01}_{-0.02}$ &  39 & $ 9.39^{+0.04}_{-0.05}$ & $ -0.64^{+0.03}_{-0.03}$ & $ 0.96^{+0.05}_{-0.05}$\\
\hline
\multirow{3}{*}{$5.0-7.0$} & 55 & $ 8.19^{+0.04}_{-0.04}$ & $ 0.93^{+0.03}_{-0.04}$ & $ 0.88^{+0.01}_{-0.01}$ &  52 & $ 8.20^{+0.04}_{-0.04}$ & $ -0.13^{+0.05}_{-0.04}$ & $ 0.58^{+0.05}_{-0.04}$\\
 &  32 & $ 8.72^{+0.04}_{-0.05}$ & $ 0.75^{+0.04}_{-0.03}$ & $ 0.93^{+0.02}_{-0.02}$ &  30 & $ 8.70^{+0.04}_{-0.03}$ & $ -0.23^{+0.04}_{-0.05}$ & $ 0.72^{+0.05}_{-0.05}$\\
 &  31 & $ 9.22^{+0.04}_{-0.03}$ & $ 0.51^{+0.02}_{-0.02}$ & $ 0.97^{+0.01}_{-0.01}$ &  30 & $ 9.23^{+0.04}_{-0.04}$ & $ -0.51^{+0.03}_{-0.03}$ & $ 0.95^{+0.04}_{-0.04}$\\
\hline
\enddata
\end{deluxetable}

We also show our line-ratio measurements on the Ne3O2 vs. RO2Ne3 diagram in the bottom panel of Figure \ref{fig:o32_r23}. Because the \neiii\ and H$\delta$ lines are faint compared to \oiii\ and \oii, the $z\sim 0$ sample on this diagram is biased toward bright, intensely SF galaxies, and thus overlaps closely with the $z\sim 2$ stacks. Though there is a large degree of scatter, the individual galaxies and stacks evolve along an overlapping locus on this diagram, with Ne3O2 increasing and RO2Ne3 decreasing with increasing redshift at fixed stellar mass. It is possible that the $z\sim 2 $ and $z\gtrsim 3.5$ stacks occupy distinct sequences in this diagram, as is the case in the \nii/H$\alpha$ BPT and O32 vs. R23 diagrams. However, it is difficult to determine how distinct the loci are due to the large measurement uncertainties.

\section{Discussion and Conclusions} \label{sec:discussion}

The trends in emission line ratios that we observe in the JADES SF galaxy sample exhibit characteristics consistent with a decreasing gas-phase metallicity with increasing redshift at fixed stellar mass \citep[e.g.,][]{2023ApJS..269...33N}. Ratios such as \sii/H$\alpha$, O32, Ne3O2, and RO2Ne3, which vary monotonically with metallicity \citep[e.g.,][]{2025arXiv250810099S}, all exhibit trends of decreasing metallicity with increasing redshift.

However, gas-phase metallicity evolution is likely not the only driver of the evolution in line ratios with redshift, because such evolution would shift galaxies along a single locus rather than perpendicular to that locus. For example, as first suggested by \citet{2025ApJ...980..242S}, the sequence of SF galaxies on the \nii/H$\alpha$ diagram sits at higher \oiii/H$\beta$ at fixed \nii/H$\alpha$ at $z\sim 3.5$ compared to $z\sim 2$, rather than evolving along a common sequence. Several works have investigated the offset between $z\sim 0$ and $z\sim 2.3$ in the \nii/H$\alpha$ diagram, finding that a harder ionizing spectrum at fixed nebular metallicity with increasing redshift provides a robust explanation for this offset \citep[e.g.,][]{2016ApJ...826..159S,2020MNRAS.495.4430T,2021MNRAS.502.2600R,2023ApJ...955...54S,2025ApJ...980..242S}. The offset between $z\sim 2$ and $z\sim 3.5$ may reflect yet a harder ionizing spectrum at $z\sim 3.5$, as the locus of SF galaxies evolves in the direction of the lower stellar metallicity Cloudy curves. However, as we can now robustly demonstrate for the first time, the loci of SF galaxies at $z\gtrsim 3.5$ overlap each other on the \nii/H$\alpha$ diagram, indicating that the ionizing conditions within SF galaxies at these redshifts at fixed nebular metallicity do not evolve strongly with time.

The loci of the SF galaxies on the \oi/H$\alpha$ diagram are at higher \oi/H$\alpha$ at fixed \oiii/H$\beta$ relative to local \hii regions and SF galaxies. Interpreting the \oi/H$\alpha$ diagram is complicated by the fact that the \oi/H$\alpha$ ratio traces photoionization, DIG, and supernova shocks \citep[e.g.,][]{2001ApJ...556..121K,2017MNRAS.466.3217Z}. DIG emission is likely negligible at the redshifts considered, given the high typical SFR surface densities at $z>2$ and the relationship between SFR surface density and DIG emission fraction at $z\sim 0$ \citep{2007ApJ...661..801O,2019ApJ...881L..35S}. Emission from gas heated by supernova shocks increases both \oi/H$\alpha$ and \oiii/H$\beta$ \citep{2001ApJ...556..121K}, thus, the positions of the stacked line ratios on this diagram are consistent with a harder ionizing spectrum and possibly strong contributions from supernova shocks.

The O32 vs. R23 diagram also exhibits a similar trend to that of the BPT diagrams, in that the $z\sim 2$ locus is offset relative to the $z\gtrsim3.5$ loci. Comparing with the Cloudy curves, we see that decreasing the stellar metallicity (thereby hardening the ionizing spectrum) shifts galaxies to the right and slightly upward on the O32 vs. R23 diagram. The bluer Ne3O2 vs. RO2Ne3 diagram displays a similar metallicity evolution trend to that seen in the O32 vs. R23 diagram. Though the stacked ratios are more scattered than in the brighter line-ratio diagrams, the Ne3O2 and RO2Ne3 ratios generally evolve in the direction of lower stellar and nebular metallicities with increasing redshift, as indicated by the Cloudy curves.

Overall, the line ratio diagrams featured in this Letter demonstrate that the ionizing spectrum does not evolve strongly at fixed nebular metallicity from $z \sim 6$ down to $z\sim 3.5$. However, by $z\sim 3.5$ (corresponding to 1.8 Gyr after the Big Bang), the shape of the ionizing spectrum begins to soften with cosmic time. The evolution in the shape of the ionizing spectrum is closely tied to changes in stellar $\alpha$/Fe abundance patterns at high vs. low redshift. It has been demonstrated that at fixed nebular oxygen abundance, stellar abundances are $\alpha$-enhanced at $z\sim 2$ relative to the solar abundance pattern, corresponding to lower inferred iron abundances at fixed nebular oxygen abundance ( e.g., \citealt{2016ApJ...826..159S,2020MNRAS.495.4430T}; \citealt{2021MNRAS.505..903C,2022ApJ...925...82K,2024ApJ...960...73C,2024MNRAS.532.3102S}) Some studies have traced the progression of $\alpha$-enhancement using other chemical element tracers such as sulfur, argon, and neon, finding similarly $\alpha$-enhanced abundance patterns up to $z\sim 5$ \citep[e.g.,][]{2024ApJ...964L..12R,2025arXiv251210130F,2025MNRAS.537.1735S,2026ApJ...997L..44R,2026MNRAS.tmp..108I}. This $\alpha$-enhancement, corresponding to a lower Fe/H at fixed O/H, results in stellar atmospheres that are more transparent to ionizing UV radiation at a given O/H, thereby hardening the ionizing spectrum. This harder ionizing spectrum at fixed nebular metallicity drives the emission-line ratio trends that we observe here. Iron is largely supplied to the ISM on $\sim$1 Gyr time scales by Type Ia supernovae, while oxygen is supplied to the ISM by core-collapse supernovae on $\sim$10 Myr time scales \citep{1986A&A...154..279M,2012PASA...29..447M}. As a result, the evolution of observed $\alpha$-enhancement provides constraints on the star-formation and chemical-evolution history of the early Universe. Previous studies analyzing stars in the Milky Way have found evidence for a plateau in $\alpha$ enhancement in old stellar populations at low Fe/H, with a subsequent decrease in $\alpha$/Fe with increasing Fe/H \citep[e.g.,][]{2020ApJ...900..179K,2024MNRAS.533.2420M}. In the context of the results presented here, we associate the common emission-line locus at $z\gtrsim 3.5$ with this plateau in $\alpha$/Fe enhancement. At lower redshifts, the evolution of the emission-line locus towards the $z\sim 0$ sequence mirrors the decrease in $\alpha$/Fe enhancement at increasing Fe/H, where Fe/H is a proxy for cosmic time (see also \citealt{2021MNRAS.505..903C}).

In addition to non-evolving abundance patterns, the non-evolving ISM conditions inferred at $z\gtrsim 3.5$ indicate that strong-line metallicity indicators calibrated to $z\sim 3.5$ SF galaxies may be applicable to SF galaxies at higher redshifts. Such a calibration enables studies of chemical abundances into the Epoch of Reionization with greatly expanded sample sizes. Such studies are key to our understanding of galaxy evolution in the Epoch of Reionization and into the current redshift frontier at $z\gtrsim 10$.

\begin{acknowledgments}
We would like to acknowledge the JADES team for their efforts in designing, executing, and making public their JWST/NIRSpec and JWST/NIRCam survey data. We also acknowledge support from NASA grants JWST-GO-01914 and JWST-GO-03833, and NSF AAG grants 2009313, 2009085, 2307622, and 2307623. This work is based on observations made with the NASA/ESA/CSA James Webb Space Telescope as well as the NASA/ESA Hubble Space Telescope. The data were obtained from the Mikulski Archive for Space Telescopes at the Space Telescope Science Institute, which is operated by the Association of Universities for Research in Astronomy, Inc., under NASA contract NAS5-03127 for JWST and NAS 5–26555 for HST. Data were also obtained from the DAWN JWST Archive maintained by the Cosmic Dawn Center. The specific observations analyzed can be accessed via \citet[][JADES]{https://doi.org/10.17909/8tdj-8n28}, \citet[][FRESCO]{https://doi.org/10.17909/gdyc-7g80}, and \citet[][JEMS]{https://doi.org/10.17909/fsc4-dt61}. This work used computational and storage services associated with the Hoffman2 Cluster which is operated by the UCLA Office of Advanced Research Computing’s Research Technology Group.
\end{acknowledgments}





%
\facilities{HST (WFC3), JWST (NIRCam, NIRSpec)}

\software{astropy \citep{2022ApJ...935..167A}, SpectRes \citep{2017arXiv170505165C}, {\sc prospector} \citep{2021ApJS..254...22J},  {\sc PyNeb \citep{2015A&A...573A..42L}}
          }





\bibliography{sample701}{}
\bibliographystyle{aasjournalv7}



\end{document}